Vapor deposition rate modifies anisotropic glassy structure of an anthracene-based organic semiconductor.

Authors: Camille Bishop,[1] Kushal Bagchi,[1] Michael F. Toney,[2] and M.D. Ediger[1]
1. University of Wisconsin – Madison, Department of Chemistry, Madison, WI, 53706 (USA)
2. University of Colorado Boulder, College of Engineering and Applied Science, Boulder, CO, 80309 (USA)

Abstract

We control the anisotropic molecular packing of vapor-deposited glasses of ABH113, a deuterated anthracene derivative with promise for future OLED materials, by changing the deposition rate and substrate temperature at which they are prepared. We find that, at substrate temperatures from $0.65T_g$ to $0.92T_g$, deposition rate significantly modifies the orientational order in the vapor-deposited glasses as characterized by X-ray scattering and birefringence. Both measures of anisotropic order can be described by a single deposition rate-substrate temperature superposition (RTS). This supports the applicability of the surface equilibration mechanism and generalizes the RTS principle from previous model systems with liquid crystalline order to non-mesogenic organic semiconductors. We find that vapor-deposited glasses of ABH113 have significantly enhanced density and thermal stability compared to their counterparts prepared by liquid-cooling. For organic semiconductors, the results of this study provide an efficient guide for using deposition rate to prepare stable glasses with controlled molecular packing.

Introduction

Organic light emitting diodes (OLEDs), first discovered in 1987,[1] are ubiquitous in current electronic display technologies.[2-4] The rapidly growing worldwide OLED industry is currently valued at USD 38 billion, and is projected to reach USD 73 billion by 2026.[5] As the roughly 30-year industry matures, novel uses, such as in biomedical devices[6] and automotive lighting, continue to expand the industry. Since OLED technologies are so widespread and frequently used, improvements in critical device parameters such as stability, energy efficiency, and device lifetime can have a significant impact on worldwide energy and resource use.

OLED devices have been extensively optimized by chemically modifying the organic semiconductors used to transport charges.[7] Anthracene derivatives are one type of molecule commonly used for OLEDs. These molecules have good electroluminescence, photoluminescence, and electrochemical properties due to their rigid ring structure and 14 $\pi$-electron aromaticity.[8-13] While some anthracenes are prone to crystallization, chemical modification is readily achieved, and the proper choice of substituents can suppress crystallization while minimally impacting (or even improving) the electronic properties.[14, 15] The use of deuterium-substituted molecules is also an important strategy to improve chemical stability and increase emitter efficiency. The chemical stability of the molecules used in OLEDs is important for maximizing lifetime of the devices, and replacing the C-H bonds in the molecules with less labile C-D bonds inhibits degradation.[16] This is a general design mechanism that does not depend upon whether deuterium substitution is directly on aromatic rings or in other parts of the molecule. Deuterium substitution in fused aromatic systems is also beneficial as this can increase the emitter lifetime by decreasing the excited state decay constant.[17, 18] For example, deuterium substitution in emitter complexes such as Ir(ppy)$_3$ and Alq$_3$ has been shown to increase current density, external quantum efficiency, and device lifetime.[19, 20]

OLEDs are typically prepared by physical vapor deposition (PVD) in vacuum, and a second approach to optimizing OLED performance is to control the anisotropic structure of the glass prepared by deposition.[21] In the PVD process, molecules are condensed in vacuum onto a temperature-controlled substrate at a controlled deposition rate. The glasses prepared by this method are often anisotropic[21-24] and, according to the surface equilibration mechanism, this is a consequence of the high mobility of organic glass surfaces.[24, 25] During vapor deposition, enhanced mobility at the surface of the glass allows the molecules to partially equilibrate towards the preferred surface structure (which is generally anisotropic) before becoming locked into the bulk by further deposition. The anisotropic packing of these glasses can enhance efficiency by improving charge mobility and directional emission.[26] Glasses prepared by this method can homogeneously incorporate multiple components with controllable molecular orientation, which is important for the host-emitter layers that are used in OLEDs.[22, 27] The increased mobility at the free surface has also been shown to prepare glasses with enhanced thermal and kinetic stability,[25, 28] which in turn increases OLED lifetimes.[29]

The effects of deposition rate and substrate temperature upon the anisotropic structure of PVD glasses can be related quantitatively using "deposition rate-substrate temperature superposition", or RTS. During a typical deposition, molecules at the surface are kinetically trapped by further deposition before they completely equilibrate to the preferred surface structure. Further equilibration during deposition can be achieved either by depositing more slowly or by depositing at a higher temperature. When RTS describes the data, the implication is that adjusting the deposition rate and the substrate temperature are equivalent paths to producing a particular structure. We have previously reported that RTS works for systems forming glasses with liquid-crystal-like packing.[30-32] OLED systems have also been deposited at various rates and $T_{sub}$, and both variables modify the anisotropic structure, though so far RTS has not been tested over a wide range of rates and substrate temperatures.[23, 33] RTS is practically important as it reduces a two dimensional parameter space to a single variable. Conceptually, demonstration of RTS supports the applicability of the surface equilibration mechanism.

In this work, we show that the anisotropic structure and molecular packing in stable glasses of ABH113, a deuterium-substituted anthracene derivative, can be successfully described by the RTS principle. We vapor-deposit glasses over a wide range of substrate temperatures ($T_{sub}$ = 0.65 - 0.92$T_g$, where $T_g$ is the glass transition temperature) and deposition rates (0.1 to 10 Å s$^{-1}$) to modify the anisotropic molecular packing. We find that the glasses prepared at these conditions have enhanced thermal stability and greater density, relative to liquid-cooled glasses. We characterize the anisotropic order in the deposited glasses by spectroscopic ellipsometry and grazing-incidence X-ray scattering and find that the two measures of structure are described by the same scaling behavior as a function rate and temperature. The deposition rate and substrate temperature can be quantitatively related by RTS. These results indicate that we can successfully control and predict glassy packing over a wide range of conditions for a typical OLED material.

Materials and Methods

*Material Characterization*

ABH113 (SFC Co., Ltd., South Korea) was used as-received. The structure of ABH113 is shown in Figure 1. The glass transition temperature $T_g$ was determined by differential scanning calorimetry (DSC, TA Instruments Q100) at scanning rates of 5 K/min, as shown in SI Figure 1. Briefly, the as-received crystalline powder was heated to 523 K, showing an endothermic melting peak at 500 K. The sample was cooled with no apparent crystallization, and the onset of the glass transition was measured during the sample's second heating to be $T_g = 370$ K. *In situ* ellipsometry thickness measurements on vapor-deposited ABH113, shown in Figure 7, are consistent with the $T_g$ determined by DSC.

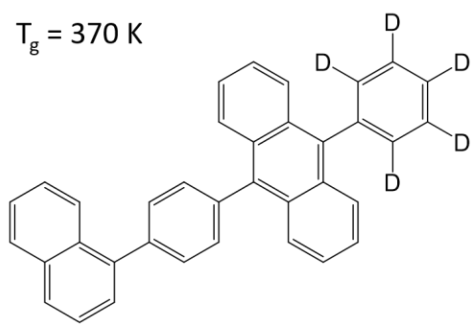

Figure 1. Molecular structure of ABH113, a deuterated anthracene derivative. $T_g$ is determined by DSC (supplemental information) and ellipsometry (shown in Figure 7).

*Vapor Deposition*

Glasses were vapor-deposited in a custom-built deposition chamber as detailed in previous publications,[24] utilizing a source-to-substrate distance of 11 cm and a base pressure of $\sim 10^{-7}$ torr. 1-inch Si <1 0 0> wafers (Virginia Semiconductor) were affixed to copper fingers using Apiezon Grease N ($T_{sub} < 293$ K) or Apiezon Grease H ($T_{sub} \geq 293$ K). The copper fingers were cooled using liquid nitrogen, and Lakeshore 336 and 340 instruments were used to control heaters to achieve the desired substrate temperature. A quartz crystal microbalance (Sycon Instruments) was used to monitor deposition rate *in situ*. Films were 100 to 250 nm thick, as characterized by ellipsometry.

*Ellipsometry measurements*

The as-deposited glasses were measured at 21 locations on each 1-inch wafer using an M-2000U spectroscopic ellipsometer from J.A. Woollam Co., Inc. Each location was measured at 7 incident angles from 45° to 75° in increments of 5°. In order to minimize errors, the ellipsometry beam was perpendicular to the sample thickness gradient (caused by the deposition geometry). Data collected at wavelengths from 500 – 1000 nm was modeled using a uniaxially anisotropic Cauchy model

$$n_z = A_z + \frac{B}{\lambda^2}, n_{xy} = A_{xy} + \frac{B}{\lambda^2}$$

with the z-direction defined as normal to the substrate. The x- and y-directions are in-plane and equivalent, i.e., the PVD glasses exhibit uniaxial anisotropy. The birefringence Δn is equal to $n_z - n_{xy}$, or $A_z - A_{xy}$, at 632 nm. Error bars shown in Figures 5 and 6 are the standard deviations of birefringence values measured at all positions of a wafer.

*X-ray scattering*

Grazing-incidence wide-angle X-ray scattering (GIWAXS) was performed at Beamline 11-3 of the Stanford Synchrotron Radiation Laboratory (SSRL) using a source-to-detector distance of 315 mm with 12.7 keV X-rays ($\lambda$ = 0.976 Å). All data shown was taken at an incidence angle of θ = 0.12°. The orientational order parameter $S_{GIWAXS}$ is defined as in equation 1:

$$S_{GIWAXS} = \frac{1}{2}(3\langle cos^2\chi\rangle - 1)$$

Where χ is the azimuthal angle in reciprocal space with χ = 0° defined along $q_z$. The $\langle cos^2\chi\rangle$ average of the scattering intensity between q values of 1.1 and 1.7 Å$^{-1}$ is defined in equation 2:

$$\langle cos^2\chi\rangle = \frac{\int_0^{90} I(\chi)(cos^2\chi)(sin\chi)d\chi}{\int_0^{90} I(\chi)(sin\chi)d\chi}$$

And is taken after an empirical B-spline background subtraction and subsequent extrapolation to all azimuthal angles, as used in previous publications[32] and shown in SI Figures 2 and 3.

*Liquid-cooled and spin-coated glass preparation*

Two reference samples were prepared for comparison with PVD glasses. One reference sample was spin-coated on a 1-inch Si wafer treated by uv/ozone cleaning. A 1.5 wt% solution of ABH113 in cyclohexanone was spin coated at 2100 rpm for 30 seconds. The sample was then baked in a nitrogen atmosphere at 80 °C for 2 minutes. This sample was 90 nm thick. As a second reference sample, a liquid-cooled glass was prepared by vapor deposition at 400 K ($T_g$ + 30 K), followed by cooling at 2 K/min.

Results

*Structure of PVD glasses of ABH113*

When deposited at various substrate temperatures ($T_{sub}$) and deposition rates, ABH113 forms glasses with a range of anisotropic packing arrangements, as shown in Figure 2. The main feature in all scattering patterns is an amorphous halo at q ≈1.4 Å$^{-1}$. This corresponds to a real-space periodicity of ≈ 4.5 Å, which we interpret as the average interaction length between two neighboring molecules. While scattering intensity at q ≈1.4 Å$^{-1}$ occurs at all azimuthal scattering angles χ, the scattering intensity is greater along $q_z$ for most deposition conditions, indicating anisotropic molecular packing.

When deposited at a constant $T_{sub}$, depositing more slowly resulted in a more isotropic distribution of scattering of the amorphous halo. The left column of Figure 2 shows a subset of samples prepared at a constant $T_{sub}$ of 275 K. At the highest deposition rate of 16 Å s$^{-1}$ (top left panel) the scattering is focused along $q_z$, indicating nearest-neighbor interactions that occur primarily out-of-plane of the film. As the deposition rate is decreased, the out-of-plane scattering decreases and at the lowest deposition rate of 0.16 Å s$^{-1}$ (bottom left panel), the scattering at q ≈ 1.4 Å$^{-1}$ is nearly equal at all scattering angles.

When deposited at a constant rate of 8 Å s$^{-1}$, increasing the $T_{sub}$ similarly makes the scattering more isotropic, as seen in the right column of Figure 2. At this deposition rate, the sample deposited at the lowest $T_{sub}$ (upper right panel) has scattering focused primarily out-of-plane. As the $T_{sub}$ of deposition is increased towards 305 K (bottom right panel), the scattering becomes isotropic at all scattering angles χ.

In Figure 2, we see that decreasing the deposition rate and increasing $T_{sub}$ have the same qualitative effect upon the anisotropic glass structure of vapor-deposited ABH113. For both parameters, more isotropic glasses are obtained when deposited at a decreased deposition rate or an increased substrate temperature. This suggests that at these conditions, molecules can more fully equilibrate at the free surface during deposition.

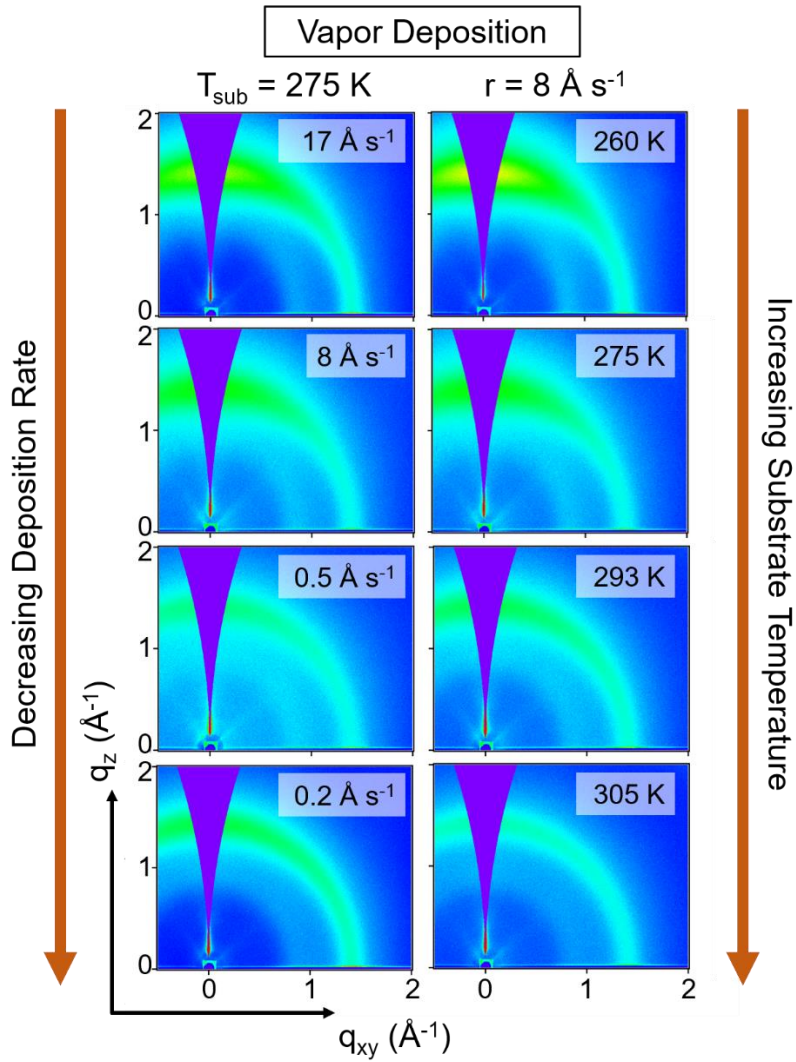

Figure 2. Grazing-incidence wide-angle X-ray scattering (GIWAXS) patterns of vapor-deposited glasses of ABH113. Left: glasses prepared at 4 deposition rates at a constant $T_{sub}$ of 275 K. Right: glasses prepared at 4 substrate temperatures at a constant rate of ~$10^{0.9}$ Å s$^{-1}$. Note that the images in the 2$^{nd}$ row of both columns of the figure are the same sample, providing a convenient frame of reference between the two control parameters.

The PVD glasses of ABH113 prepared here are distinct from those that can be prepared by conventional methods. Figure 3 shows the X-ray scattering from glasses prepared by liquid-cooling and spin-coating. For the liquid-cooled sample, the amorphous scattering peak at q ≈ 1.4 Å$^{-1}$ is completely isotropic, with equal scattering intensity at all azimuthal angles. It closely resembles the scattering patterns from the glasses vapor-deposited at the highest $T_{sub}$ and lowest rates. The spin-

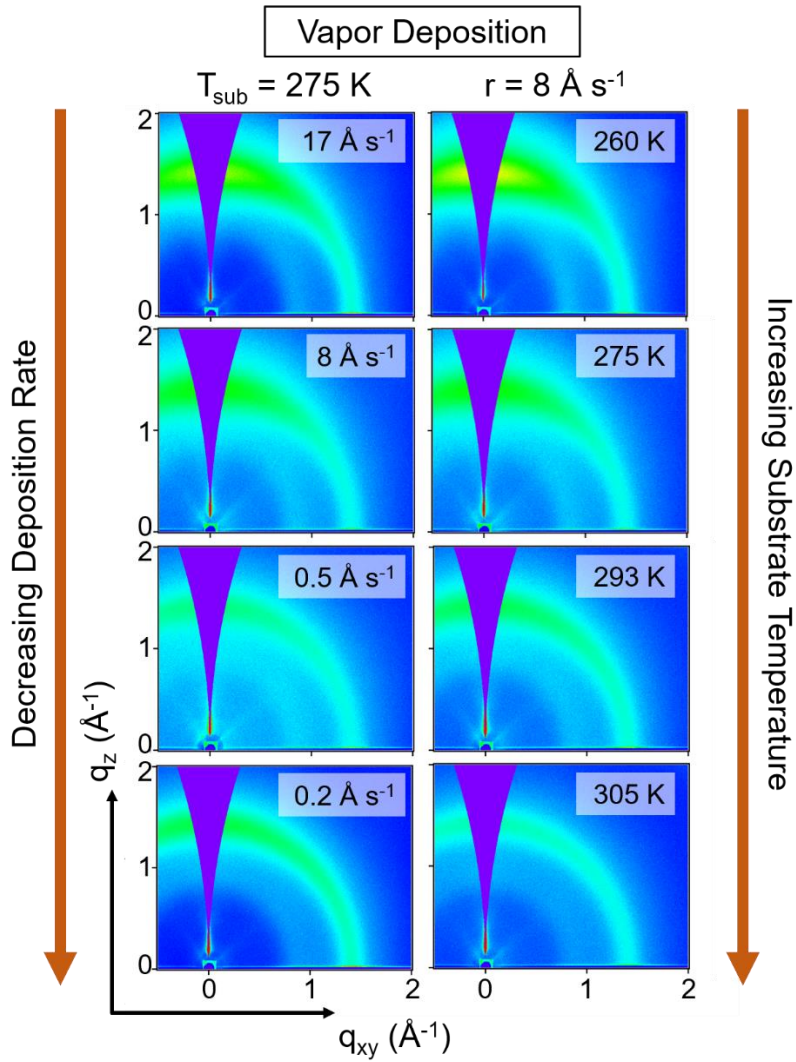

Figure 2. Grazing-incidence wide-angle X-ray scattering (GIWAXS) patterns of vapor-deposited glasses of ABH113. Left: glasses prepared at 4 deposition rates at a constant $T_{sub}$ of 275 K. Right: glasses prepared at 4 substrate temperatures at a constant rate of ~$10^{0.9}$ Å s$^{-1}$. Note that the images in the 2$^{nd}$ row of both columns of the figure are the same sample, providing a convenient frame of reference between the two control parameters.

The PVD glasses of ABH113 prepared here are distinct from those that can be prepared by conventional methods. Figure 3 shows the X-ray scattering from glasses prepared by liquid-cooling and spin-coating. For the liquid-cooled sample, the amorphous scattering peak at q ≈ 1.4 Å$^{-1}$ is completely isotropic, with equal scattering intensity at all azimuthal angles. It closely resembles the scattering patterns from the glasses vapor-deposited at the highest $T_{sub}$ and lowest rates. The spin-coated sample shows weaker scattering, though the major scattering feature is similarly isotropic. In addition, the spin-coated sample shows minor scattering features (at q ~ 1.5 Å$^{-1}$ and χ ~ 70°) whose origin is unknown.

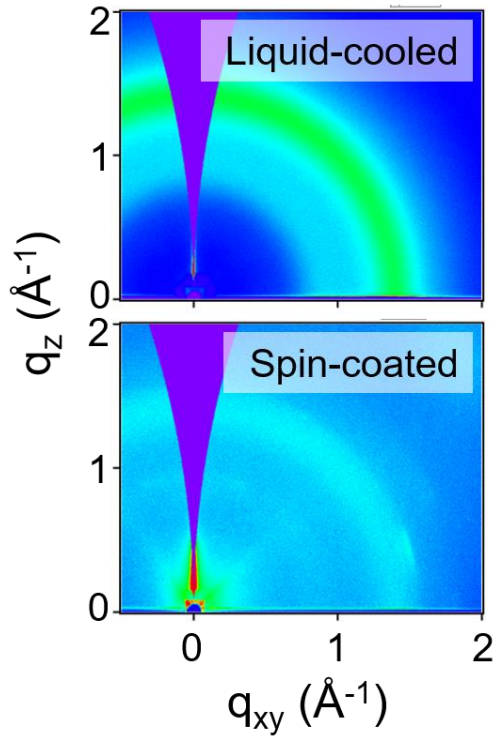

Figure 3. ABH113 glasses prepared by cooling from the equilibrium liquid and by spin-coating. For both reference glasses, scattering near q ≈ 1.4 Å$^{-1}$ is nearly isotropic.

*Test of Rate-Temperature Superposition*

The anisotropy of the X-ray scattering near q ≈ 1.4 Å$^{-1}$ can be quantified using the Hermans orientation order parameter, $S_{GIWAXS}$.[34, 35] A value of $S_{GIWAXS}$ = +1.0 corresponds to scattering focused entirely out-of-plane, while an $S_{GIWAXS}$ = -0.5 corresponds to scattering completely in-plane (along $q_{xy}$). A glass with isotropic scattering will have an $S_{GIWAXS}$ of 0. Figure 4A shows how $S_{GIWAXS}$ varies as a function of the log of the deposition rate, for several $T_{sub}$ values. At each substrate temperature, $S_{GIWAXS}$ becomes moderately more positive as the deposition rate is increased, corresponding with an increase in out-of-plane scattering. By examining Figure 4A at a single deposition rate, we can also observe the variation of scattering anisotropy with substrate temperature. Samples were deposited at a rate near 8 Å s$^{-1}$ for all nine $T_{sub}$ studied. The samples deposited at the lowest $T_{sub}$ have mildly negative or zero values of $S_{GIWAXS}$, indicating a slight tendency for in-plane scattering intensity or completely isotropic glasses. As $T_{sub}$ is lowered, the values for $S_{GIWAXS}$ become more positive (increase in out-of-plane scattering), similarly to the case in which deposition rate is lowered.

In Figure 4B, we examine the relationship between $T_{sub}$ and deposition rate further by attempting a quantitative rate-temperature superposition. The scattering anisotropy depends upon both the deposition rate and the substrate temperature, and we combine the two parameters into a single "effective deposition rate" rate, i.e., the rate at which the glass must be deposited at the reference temperature in order to achieve the specified scattering anisotropy. We choose room temperature (293 K) as the reference temperature, as this is often the temperature at which glasses are prepared for OLEDs on an industrial scale. We test a multitude of "shift factors" to find a smooth collapse of data into a master curve, as illustrated graphically in previous publications.[31, 32] For each tested shift factor, we transform the actual deposition rate to an effective deposition rate based upon the $T_{sub}$; glasses deposited at lower $T_{sub}$ will have lower effective rates than actual rates. A shift factor of 17 K per decade results in the successful "master curve" shown in Figure 4B. This collapse means that raising $T_{sub}$ by 17 K during deposition will prepare a glass with $S_{GIWAXS}$ equivalent to one made by depositing 10 times more slowly at the original substrate temperature. We note that at the lowest effective deposition rates, $S_{GIWAXS}$ approaches zero (the glass becomes more isotropic) after reaching a minimum. This is consistent with a mechanism proposed for non-liquid crystal molecules, in which $T_{sub}$ and rate control not only the rate of equilibration, but also the equilibration depth.[32]

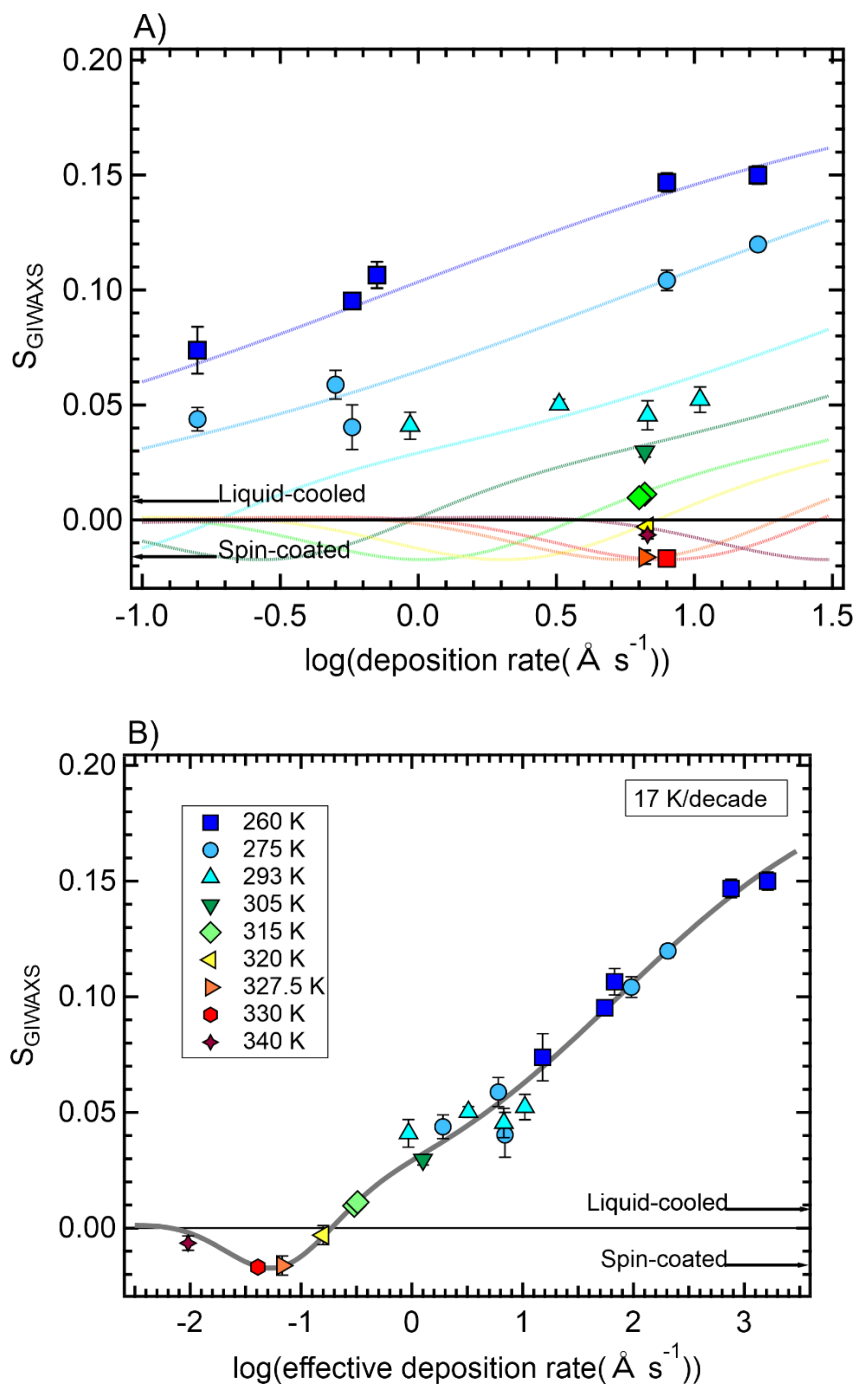

Figure 4. The orientational Hermans order parameter $S_{GIWAXS}$ versus the actual and effective deposition rate. A) $S_{GIWAXS}$ of the q ~ 1.4 Å$^{-1}$ peak as a function of the log of the deposition rate for nine substrate temperatures. Error bars given are the standard deviation in calculated $S_{GIWAXS}$ over three spots on each sample. Generally, $S_{GIWAXS}$ decreases (less out-of-plane scattering intensity) with increasing $T_{sub}$ and decreasing deposition rate. B) $S_{GIWAXS}$ versus an effective deposition rate at 293 K calculated by applying the rate-temperature superposition (RTS) principle using the "shift factor" in which raising $T_{sub}$ by 17 K during deposition has the equivalent effect on $S_{GIWAXS}$ as lowering the deposition rate by a factor of 10. As a guide to the eye, we empirically fit a curve to the data in panel B; this same curve is horizontally transformed by the shift factor and used in panel A.

We also measured the optical birefringence of the PVD glasses of ABH113. The birefringence is the difference between the out-of-plane and in-plane indices of refraction; a positive birefringence indicates a higher refractive index out-of-the-plane of the film. The birefringence is primarily controlled by the average molecular orientation,[24] with a positive birefringence indicating that the molecular axis with the largest polarizability tends to be vertically oriented in the film. For ABH113, the molecular long axis (along the 9,10 axis of the anthracene) is expected to have the largest polarizability. Figure 5A shows the birefringence of glasses prepared at several $T_{sub}$ as a function of the deposition rate. Generally, when deposited at higher rates, the birefringence of the glasses becomes more negative, i.e., the long axes of the ABH113 molecules tend to lie in the plane of the substrate.

Despite the noise in the birefringence data (due to thin samples), a master curve produced by rate-temperature superposition results in a clear trend. Figure 5B shows the birefringence as a function of the effective deposition rate at 293 K. For this curve, we used the "shift factor" found for the $S_{GIWAXS}$ data (17 K / decade) and found that it superposes the data effectively. Therefore, we determine that all data is consistent with one shift parameter. This differs from other systems that have been previously studied, which show different shift factors for the scattering anisotropy and the optical birefringence; this topic is further explored in the discussion.

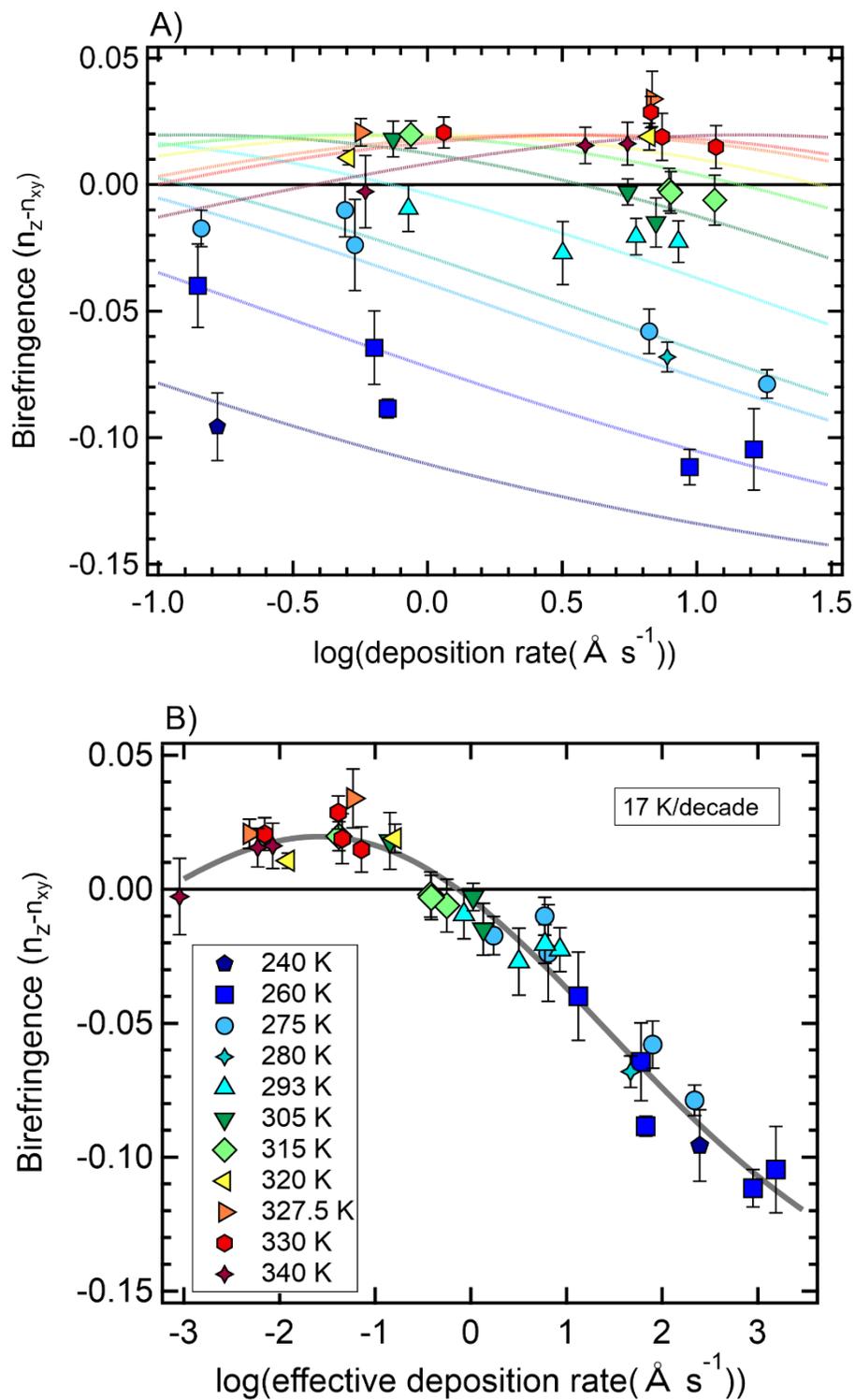

Figure 5. The birefringence, a measure of molecular orientation, versus the actual and effective deposition rate. A) Birefringence, the difference in out-of-plane and in-plane refractive index, versus the log of the deposition rate for eleven substrate temperatures. Generally, birefringence

becomes more positive with increasing $T_{sub}$ and decreasing deposition rate. B) Birefringence versus an effective deposition rate at 293 K calculated by applying the rate-temperature superposition (RTS) principle using a shift factor of 17 K/decade. The curve fit to the data in panel B is also used in panel A.

Figures 4 and 5 suggest a strong correlation between anisotropic order in the x-ray scattering and the molecular orientation. We test this directly in Figure 6A, by plotting $S_{GIWAXS}$ vs. the birefringence. The two observables correlate well, with an $R^2$ for a linear fit of 0.97, indicating that these two types of anisotropy likely share a common origin, which is the anisotropic structure of the liquid surface according to the surface equilibration mechanism. The negative slope of the data in Figure 6A can be rationalized as follows: At the left side of the panel (lowest $T_{sub}$ and highest deposition rates), molecules have a strong tendency to lie in the plane of the substrate (as indicated by negative birefringence) which facilitates a tendency towards face-on stacking in the z direction (which concentrates scattering along the z axis, leading to a positive value of $S_{GIWAXS}$.) The best-fit line in Figure 6A has a non-zero y-intercept, that is, when the birefringence is equal to zero, $S_{GIWAXS}$ is equal to 0.024. While this does not match simple expectations, there is precedent for this mismatch.[30] It is important to emphasize that a glass with a birefringence of zero (or a $S_{GIWAXS}$ value of zero) is necessary, but not sufficient, condition for a truly isotropic sample.

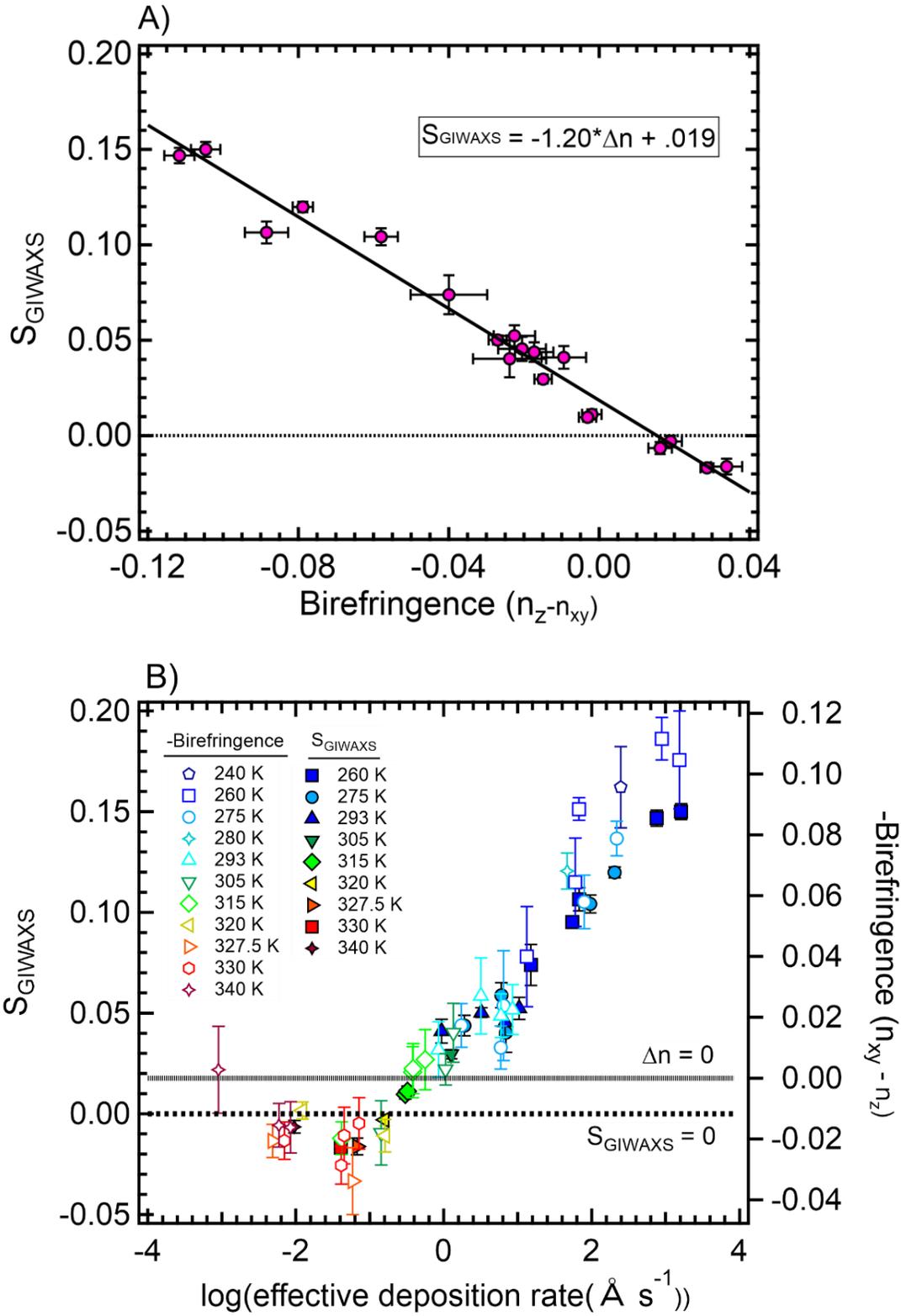

Figure 6. S$_{GIWAXS}$ and birefringence, two measures of anisotropic order, show a strong correlation that allows them to be related by the same rate-temperature superposition shift factor. A) S$_{GIWAXS}$

vs. birefringence of the subset of samples measured by both GIWAXS and ellipsometry. A linear fit is shown, with an $R^2$ of 0.97. B) $S_{GIWAXS}$ and the negative of the birefringence, plotted against the log of the effective rate at 293 K shows that the two measures of order follow the same trend. Filled symbols correspond to the X-ray data, while open symbols show the birefringence.

Figure 6B shows that the birefringence and $S_{GIWAXS}$ values can be combined in a single superposition plot. Within error, the two follow the same functional trend. We note the two different y-axis positions for the "zero" in Figure 6B; this reflects the non-zero y-intercept in Figure 6A.

*Thermal Stability of Vapor-Deposited Glass*

We used ellipsometry measurements during heating to show that the vapor-deposited ABH113 glasses have characteristics of an "ultrastable glass" that include enhanced thermal stability and greater density compared to the liquid-cooled glass. To illustrate this, Figure 7A shows ellipsometry measurements during temperature cycling for a glass that was vapor-deposited at 6 Å s$^{-1}$ at $T_{sub}$ = 305 K. The sample thickness was monitored during heating and subsequent cooling, all at 1 K/min. During the first heating experiment, the glass gradually expands as a solid, as a result of thermal expansion. This gradual expansion continues even after the ordinary glass transition temperature $T_g$ (370 K) is reached. At the onset temperature ($T_{onset}$) of about 384 K, the film thickness rapidly expands due to the transformation of the as-deposited glass into the less dense supercooled liquid. After the transformation into the supercooled liquid was completed, the sample was cooled at 1 K/min, and vitrified into an ordinary liquid-cooled glass at $T_g$. The ordinary glass was then heated and cooled two more times, with good reproducibility.

The as-deposited glass has enhanced thermal stability, as evidenced by the high value of $T_{onset}$ in the temperature ramping experiment. $T_{onset}$ of the as-deposited glass is measured upon the first heat to be 384 K. $T_g$ is determined from the three cooling experiments to be 370 K, which is consistent with the $T_g$ measured by DSC as shown in SI Section 1. The $T_{onset}$ for the as-deposited glass of 384 K is 14 K above $T_g$, equal to 1.04$T_g$. $T_g$ and $T_{onset}$ are determined from the intersection of the tangents drawn to the thermal expansion of the glass and supercooled liquid. Typical values for "ultrastable glasses" of organic molecules range from ~1.02$T_g$ to 1.06$T_g$,[36] showing that stability of vapor-deposited ABH113 is comparable to other stable vapor-deposited glasses, including previously studied organic semiconductors.[24]

The temperature ramping experiments in Figure 7A also show that the as-deposited glass is denser than the liquid-cooled glass. At room temperature, the as-deposited glass is 3.3 nm thinner than the liquid-cooled glass - a 1.7% increase in density. This value is similar to the largest density increases measured for PVD glasses of other organic semiconductors[24] and similar molecules.[37, 38] Both the higher density and the increased thermal stability of the vapor-deposited glass can be explained by the ability of molecules to find packing arrangements closer to equilibrium during deposition below $T_g$, as described further in the discussion.

To quantitatively evaluate the thermal stability of PVD glasses of ABH113, we performed isothermal annealing experiments above $T_g$. Experiments were performed on a glass deposited at 6 Å s$^{-1}$ at $T_{sub}$ = 305 K. In an isothermal annealing experiment, as shown in Figure 7B, the as-deposited glass is held at a single temperature above $T_g$ and monitored as it transforms to the supercooled liquid. To quantify the thermal stability, the overall time of the transformation is compared to the structural relaxation time of the supercooled liquid at the annealing temperature. For this experiment, we held the as-deposited glass at 385 K, 15 K above $T_g$. We estimate the structural relaxation time $\tau_\alpha$ at 385 K to be 0.09 seconds using temperature jump experiments, shown in detail in Figures S4 through S6. When the as-deposited glass is held at 385 K, after a short induction period, the overall measured film thickness slowly begins to increase as the glass transforms into a supercooled liquid. We model the film as a single layer; while the supercooled liquid and as-deposited glass have slightly different optical constants, the overall film thickness is satisfactorily modeled. We assume, consistent with previous literature results, that the glass transforms to a supercooled liquid through a constant-velocity transformation front that initiates at the surface, increasing the overall film thickness in a linear fashion.[39-41] After 3067 seconds, the film reaches a plateau thickness, indicating that it is completely transformed to the supercooled liquid. Given our estimate of the structural relaxation time $\tau_\alpha$, the transformation occurs in approximately 34,000 $\tau_\alpha$, indicating a highly stable glass. (See SI pages 3-5 for details.) For comparison, a liquid-cooled glass transforms in approximately 1 $\tau_\alpha$, showing that the vapor-deposited glass shown here has enhanced stability.

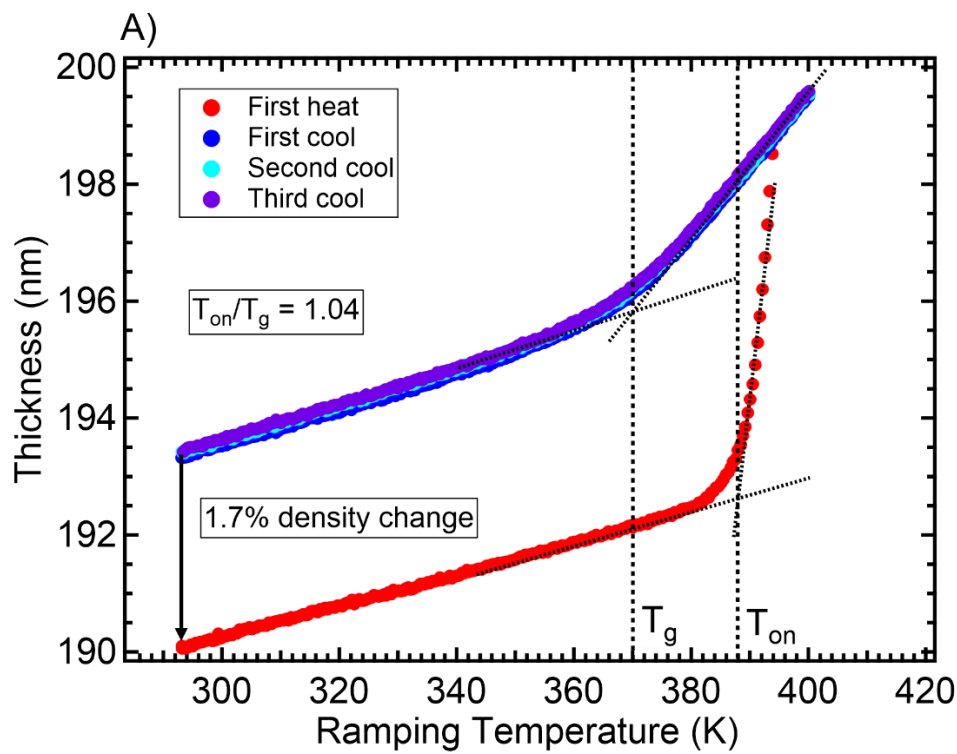

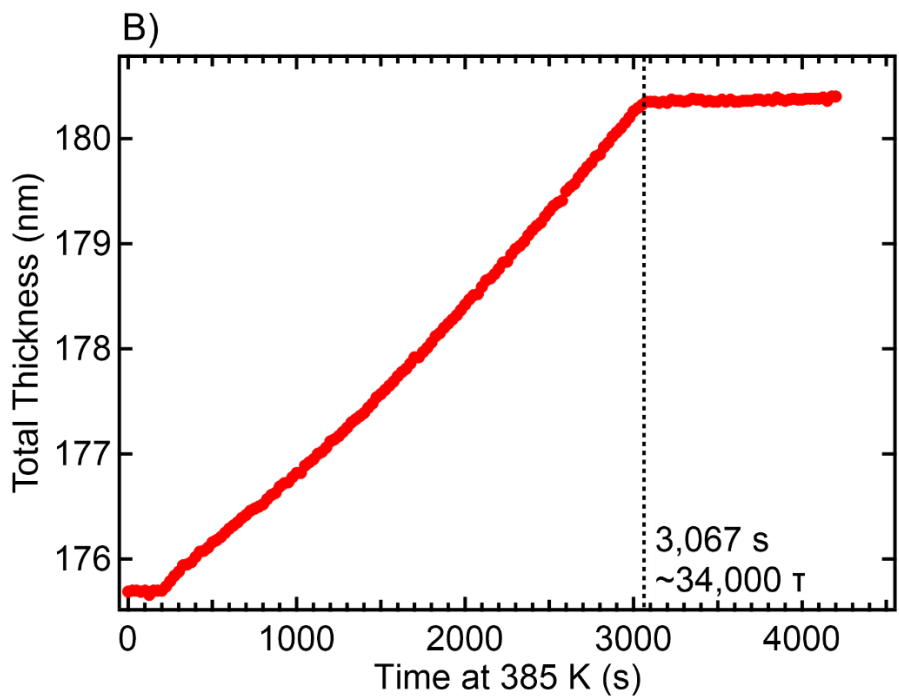

Figure 7. Ellipsometry measurements during sample heating show that a vapor-deposited glass of ABH113 has high thermal stability and high density. A) Thickness vs. temperature of a vapor-deposited glass of ABH113 when heated and cooled at 1 K/minute. The glass was deposited at 6 Å s$^{-1}$ at $T_{sub}$ = 305 K. The temperature at which the as-deposited glass transforms to the liquid, $T_{onset}$, is equal to 384 K, 14 K above the glass transition temperature $T_g$, showing enhanced thermal stability. B) Isothermal transformation of a glass held at 385 K, 15 K above $T_g$. The glass transforms in 3067 seconds, which is equal to $34,000\tau_\alpha$.

Discussion

In many respects the structure and stability of PVD glasses of ABH113 are similar to previously-studied non-mesogenic organic semiconductors with relevance for OLEDs. The birefringence and $S_{GIWAXS}$ as a function of substrate temperature are qualitatively similar to those previously reported for TPD, NPD, and DSA-Ph.[24] In particular, for these three systems and ABH113, glasses deposited at low substrate temperatures have the molecular long axis oriented towards the plane of the substrate, on average,[24, 42] while a slight tendency towards vertical orientation is observed for deposition at substrate temperatures close to $T_g$. As noted above, the density and thermal stability of PVD glasses of ABH113 are typical of the most stable glasses prepared to date – and in this way they are similar to other organic semiconductors that have been studied. The similarity of the PVD glasses of ABH113 to those of other organic semiconductors makes ABH113 a suitable candidate for testing deposition rate-substrate temperature superposition. We show below that the superposition works well for ABH113 and we expect that it will also apply to many other organic semiconductors.

*Generality of Rate-Temperature Superposition*

For some systems, the effects of deposition rate and substrate temperature upon the anisotropic structure of PVD glasses can be related quantitatively using "deposition rate-substrate temperature superposition", or RTS. In PVD, molecules at the surface attempt to equilibrate towards the equilibrium liquid surface structure; this happens relatively rapidly given that mobility at the surface is many orders of magnitude faster than mobility in the bulk.[43] Typically, however, these molecules are kinetically trapped by further deposition before they completely equilibrate. Further equilibration during deposition can be achieved either by depositing at a higher temperature or depositing more slowly. When RTS describes the data, it suggests that adjusting the deposition rate and the substrate temperature are equivalent paths to producing a particular structure. Demonstration of RTS is consistent with the surface equilibration mechanism. On a practical level, RTS reduces a two-dimensional parameter space to a single combined variable that controls glass anisotropy. This is important as the anisotropy of vapor-deposited organic semiconductors has been shown to significantly impact device performance, through enhancing charge mobility and outcoupling efficiency.[21]

We have previously reported that RTS works for three systems that form glasses with liquid-crystal-like packing.[30-32] RTS has been shown to work for itraconazole and posaconazole, model systems of pharmaceutical importance.[31, 32] PVD glasses of both of these systems can exhibit high levels of vertical orientation and smectic-like layering. RTS has also been demonstrated for a phenanthroperylene ester that is an organic semiconductor.[30] This phenanthroperylene ester is also a liquid crystal mesogen, forming hexagonal columnar structures both in equilibrium and via PVD.

The demonstration here that PVD glasses of ABH113 follow RTS extends previously published work in several ways. Most importantly, ABH113 is more typical of the molecules used in OLED devices, including deuterium-substitution to increase chemical stability and emitter efficiency. PVD glasses of ABH113 do not show strong indications of liquid-crystalline-like packing. In this way, ABH113 is different than the three model systems discussed in the previous paragraph and more similar to common OLED molecules.[24, 33, 42] One previous study on Alq3, a common OLED emitter, used two deposition rates and found that translational layering order may follow an RTS.[33] This previous work, together with the more comprehensive study of ABH113 presented here, suggests that the RTS principle is broadly useful to predict and control the structure of glasses of industrially relevant molecules used in OLED devices. A second important aspect of the current work is that RTS was observed to work with the same shift factor for both x-ray and birefringence measurements. More complex behavior has been observed for itraconazole and the phenanthroperylene ester, and this complexity is likely associated with the longer range order present in PVD glasses of these molecules.[30, 31] A third important aspect of our work on ABH113 is the wide substrate temperature range over which RTS has been shown to work, extending to substrate temperatures as low as $0.65T_g$ for the birefringence measurements. This temperature range is wider than the range tested in previous work and includes room temperature. The demonstration that RTS works at even very low substrate temperature is especially relevant for OLED molecules, which are often deposited at room temperature and often have high glass transition temperatures. Previous work by Yokoyama and coworkers can be interpreted in retrospect as being consistent with RTS, as they found that the molecular orientation of several vapor-deposited OLED molecules depended on deposition rate when deposited at room temperature (~0.70 $T_g$).[21, 23] Finally, we have shown the ABH113 glasses that follow RTS are also highly stable and higher in density than a liquid-cooled glass. High thermal stability has been shown to extend device lifetimes.[29, 44]

The RTS shift factor is a measure of how much the final structure of the glass depends upon both the $T_{sub}$ and the rate. The RTS shift factor of 17 K/decade found for ABH113 in this work indicates that, at a single substrate temperature, the molecular packing can be adjusted considerably by changing the deposition rate. This could be important in the manufacture of OLEDs where deposition chambers typically only allow room temperature depositions.

*Surface Relaxation Inferred from Rate-Temperature Superposition*

The observation of RTS in PVD glasses of ABH113 is consistent with the surface equilibration mechanism and can be used to gain insight into the surface relaxation process that is responsible for the as-deposited glass structure. From Figure 6B, we infer that the surface relaxation time that controls PVD glass structure is on the order of $10^{2.5}$ s at 293 K (using the effective rate associated with the minimum in the figure and a 1 nm estimate of the relevant surface layer). From this perspective, we see that PVD is useful for preparing glasses with a range of structures because the relevant (surface) relaxation time is not so different from the time required to deposit one molecular layer, even more than 70 K below $T_g$. For comparison, the corresponding bulk relaxation time ($\tau_\alpha$) at 370 K is found to be about $10^4$ s (see SI) and by extrapolation would be about $10^{29}$ s at 293 K. While this extrapolation is certainly rough, it serves to make the point that the equilibrium relaxation times for ABH113 in the bulk are extremely long. Therefore, once a particular anisotropic glass of ABH113 is prepared, it is expected to maintain this structure for extremely long times in the glassy state.

It is useful to compare the surface relaxation process for ABH113 with other systems studied by PVD. In contrast to previously studied systems, the metrics of anisotropic structure for ABH113 glasses (birefringence and $S_{GIWAXS}$) can be described with a common shift factor. The shift factor of 17 K/decade allows us to calculate the activation energy for the molecular rearrangements responsible for anisotropic order in ABH113 glasses and we find a value of approximately 100 kJ/mol. [For reference, the activation energy for the corresponding bulk relaxation process ($\tau_\alpha$) at $T_g$ is much larger, about 900 kJ/mol (see SI).] For itraconazole and posaconazole, which form layered smectic-like glasses via PVD, the shift factor associated with molecular reorientation is different from the shift factor associated with translational order.[31, 32] For these two systems, the activation energy for the process controlling molecular reorientation is about 390 kJ/mol.[32] Recent work by Li et al.[45] provides a way to understand the much larger activation energies reported for posaconazole and itraconazole. These authors show that for systems without hydrogen bonding, surface diffusion coefficients are strongly influenced by the depth penetrated by surface molecules. Since posaconazole and itraconazole are nearly vertical at the surface of the equilibrium liquid, they penetrate further from the free surface than is expected for ABH113; the larger penetration depth leads to both slower surface diffusion and a larger activation energy for surface diffusion. Activation energies for the translational layering shift factor for itraconazole are even larger (920 kJ/mol) and this is consistent with the idea that achieving this type of order requires mobility even further away from the free surface. The rather low activation energy for the orientational ordering of ABH113 found in this work suggests that the process takes place at only the very top (1-2 nm) of the film during deposition.[30] The activation energy for ABH113 reported here is similar to the value reported for the surface reorientation process of a phenanthroperylene ester, which has hexagonal columnar liquid crystal phases.[30]

As a final comment, we note that the surface equilibration process that gives rise to anisotropic structures in PVD glasses need not be the same as surface diffusion. The two processes may be related, consistent with the reported correlation between glass stability and surface diffusion coefficients for several molecular glasses.[46] On the other hand, Samanta et al.[47] have shown that stable glasses can be formed for some molecules even when surface diffusion is negligibly slow.

Conclusions

We have shown that the anisotropic structure of vapor-deposited glasses of an organic semiconductor is consistent with the deposition rate-substrate temperature superposition principle. Over a wide range of substrate temperature, from $0.65 - 0.92$ $T_g$, increasing the substrate temperature has the same effect on the glass structure as decreasing the deposition rate. Notably, these conditions include room temperature (293 K), and thus our work enables structure manipulation via deposition rate in vacuum depositors used industrially. The existence of a single shift factor that describes RTS for both X-ray and optical birefringence experiments implies a particularly simple formation route, consistent with the surface equilibration mechanism. Glasses of ABH113 also show enhanced thermal stability and density. Thus, for this typical OLED molecule, glass packing can be efficiently manipulated over a large range of structures, through predictable changes in either substrate temperature or deposition rate, while maintaining the tight packing associated with long-lived and energy-efficient devices.

Acknowledgements

We gratefully acknowledge LG Chem for support of this research. Additional support was received from NSF through the University of Wisconsin Materials Research Science and Engineering Center (Grant DMR-1720415). Use of the Stanford Synchrotron Radiation Lightsource, SLAC National Accelerator Laboratory, is supported by the US Department of Energy, Office of Science, Office of Basic Energy Sciences under Contract DE-AC02-76SF00515. C.E.B. acknowledges support from the National Research Council Research Associateship Program. We thank Zahra Fakhraai, Jaritza Gómez, and Diane Walters for performing preliminary experiments that motivated the present work, as well as Zahra Fakhraai and Lexi Zhang for helpful discussions.

Data Availability Statement

The data that support the findings of this study are available from the corresponding author upon reasonable request.

Supporting Information for "Deposition rate modifies anisotropic structure of vapor-deposited stable glasses of a deuterium-substituted organic semiconductor"
Authors: Camille Bishop, Kushal Bagchi, Michael F. Toney, and M.D. Ediger

Determination of Glass Transition Temperature

Glass transition temperature was determined using differential scanning calorimetry. The as-received crystalline material was heated above its melting point and cooled at 5 K/min to prepare a bulk liquid-cooled glass. Upon heating, a glass transition temperature at 370 K corresponds to that of the liquid-cooled glass.

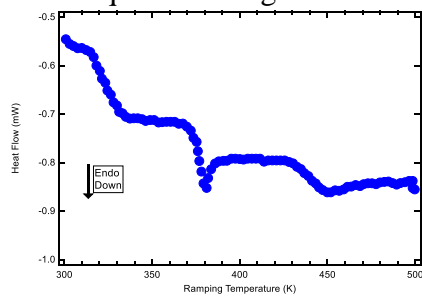

Figure S1. Raw heating curves for DSC scans of bulk ABH113 glass. The bulk glass was prepared by cooling the liquid at 5 K/min from 510 K. It was then heated at 5 K/min to investigate phase transitions of the bulk glass.

Background Subtraction and Extrapolation for $S_{GIWAXS}$

Values of $S_{GIWAXS}$ shown in Figures 4 and 6 of the main text are calculated according to equations 1 and 2 in the main text, repeated here for reference:

$$S_{GIWAXS} = \frac{1}{2}(3\langle cos^2\chi \rangle - 1)$$

where

$$\langle cos^2\chi \rangle = \frac{\int_0^{90} I(\chi)(cos^2\chi)(sin\chi)d\chi}{\int_0^{90} I(\chi)(sin\chi)d\chi}$$

Data is first integrated and background subtracted, as shown in Figure S2.
The raw integrated intensity is taken between q = 1.2 and 1.65 Å$^{-1}$, low-q background between q = 0.975 and 1.2 Å$^{-1}$, and high-q background between q = 1.65 and 1.875 Å$^{-1}$. The sharp decrease at χ ≈ 0° to 10° is due to the "missing wedge" caused by detector geometry. The sharp intensity change at χ ≈ 88° is due to edge effects from the detector. Both regions are not used in subsequent fitting.

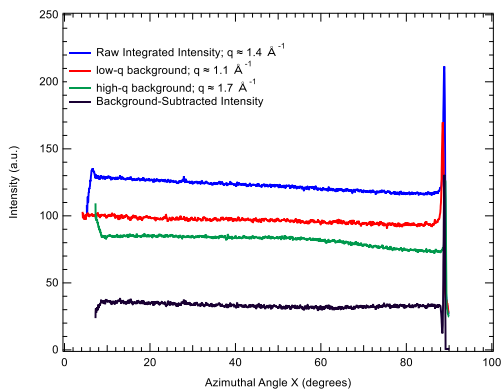

Figure S2. I vs. χ curves used for background subtraction for glass prepared at $T_{sub}$ = 315 K, rate = 6 Å s$^{-1}$.

An empirical B-spline is used to fit the data and fill in the intensity from χ ≈ 0° to 10° and χ > 88°, as shown in Figure S3. Fit was extrapolated to χ = 0° to 90° to determine the $\langle cos^2 \chi \rangle$ average. Several polynomial and linear fits were tried; all resulted in the same values of S$_{GIWAXS}$ +/- 0.002.

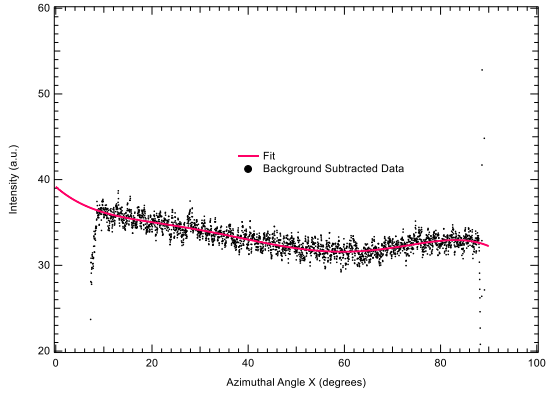

Figure S3. B-spline fit to the background-subtracted data to fill in missing intensity.

Estimating Bulk Relaxation of ABH113

To quantify the thermal stability of the as-deposited glass shown in Figure 7b in the main text, we must determine the structural relaxation time. We determine the structural relaxation time using temperature jump ellipsometry experiments, in which the film thickness is perturbed by a sudden temperature jump, and the time required for the film to equilibrate to a constant thickness at that temperature is measured. An example temperature protocol, with the resulting thickness changes, is shown in Figure S4. The data is smoothed using a Savitsky-Golay algorithm, and is fit to a KWW exponential function of the form $h = A * exp\left(-\left(\frac{t}{\tau}\right)^\beta\right) + h_0$, shown in Figure S5. The initial and final film thicknesses are fixed, and the data is fit to find the equilibrium structural relaxation time τ. The resulting $\tau_\alpha$ values at several temperatures above $T_g$ are shown in Figure S6. Because of the insufficient time resolution required to measure $\tau_\alpha$ at 385 K, the annealing temperature shown in Figure 7B, measurements are made at lower temperatures and then extrapolated to obtain $\tau_\alpha$ at 385 K.

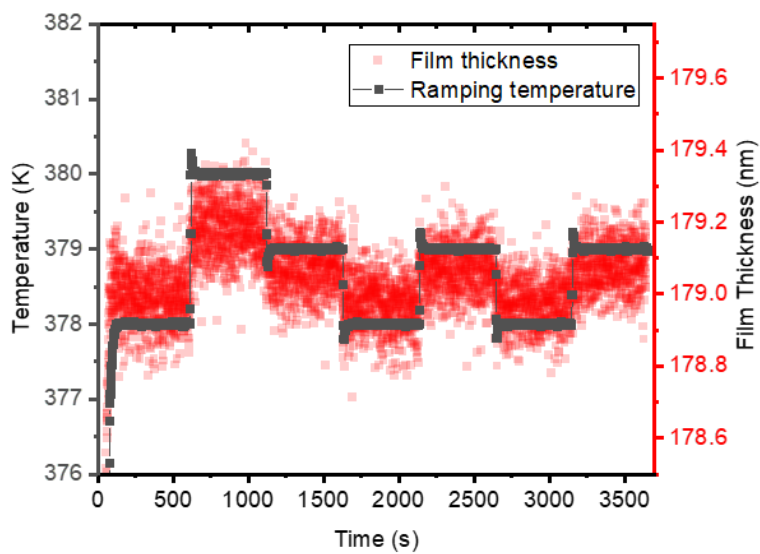

Figure S4. Example temperature protocol with time (black points, left y-axis) and film thickness response (red points, right y-axis) of a liquid-cooled glass of ABH113.

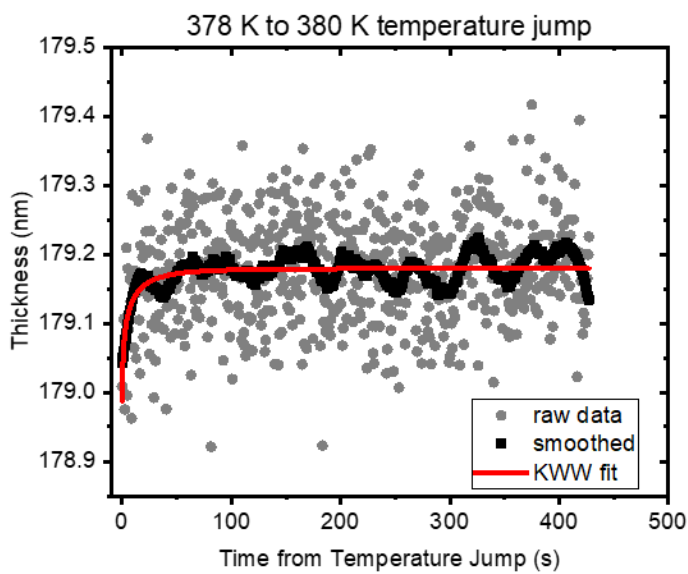

Figure S5. Example thickness response of the liquid-cooled glass to the first temperature jump shown in Figure S4. Smoothed data is shown as a guide to the eye; KWW function is fit to the raw data, with the initial and final film thicknesses set to the average thickness of the glass at the initial and final temperatures, respectively.

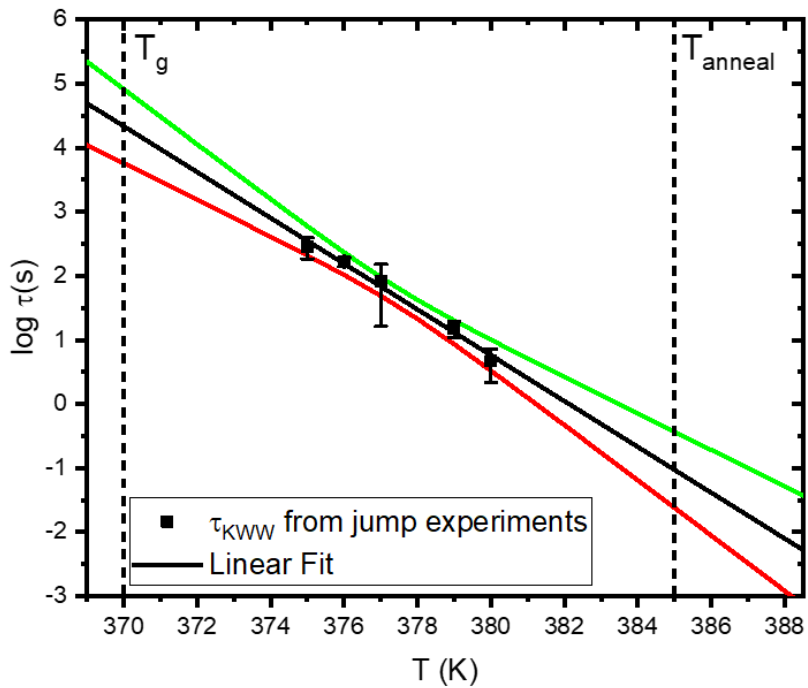

Figure S6. $\tau_{KWW}$ values determined by several experiments similar to that shown in Figure S3. Error bars shown are standard deviations in $\tau$ values from three experiments each. A linear fit with 95% confidence intervals is shown.

     The linear fit can be used to estimate the $\tau$ value at 385 K, the annealing temperature shown for the thermal stability measurements shown in Figure 7 of the main text. The linear fit yields $\tau$ = 0.09 s, with the upper and lower confidence bounds yielding $\tau$ = 0.370 s and $\tau$ = 0.024 s, respectively. We assume the value of the bulk relaxation time $\tau_\alpha$ to be equal to this $\tau$ value.

     The transformation time $t_{transform}$ for the glass shown in Figure 7 of the main text is 3067 seconds. The quantity $t_{transform}/\tau_\alpha$ is a measure of how long the glass takes to transform proportionally to the structural relaxation time. Given the $\tau$ values found above for the linear fit, upper, and lower confidence bounds, the glass takes approximately 34,000 structural relaxation times to transform.